\documentclass[11pt]{article}

\usepackage[margin=1in]{geometry}
\usepackage{times}
\usepackage{hyperref}
\usepackage{url}
\usepackage{natbib}
\usepackage{amsmath}
\usepackage{graphicx}
\usepackage{booktabs}
\usepackage{enumitem}
\usepackage{xcolor}
\usepackage{microtype}
\usepackage[T1]{fontenc}
\usepackage[utf8]{inputenc}

\hypersetup{
    colorlinks=true,
    linkcolor=blue,
    citecolor=blue,
    urlcolor=blue
}

\title{\textbf{The Context Access Divide: Interaction-Level Architecture\\
as a Complementary Dimension of Agentic Inequality}}

\author{
    Masahiro Fujita\\
    Faculty of Sociology, Kansai University\\
    \texttt{m.fujita@kansai-u.ac.jp}
}

\date{July 2026\\[0.5em]
\small Preprint --- submitted to arXiv cs.CY}

\begin{document}

\maketitle

\begin{abstract}
\citet{Sharp2025} introduce ``agentic inequality'' as a framework for analyzing disparities in access to AI agents across three dimensions: availability, quality, and quantity. These dimensions operate at the level of persons and organizations, characterizing who can access agents and at what capability. This paper identifies a structurally important divide that operates at a different level of analysis---the individual interaction---and that the \citet{Sharp2025} framework does not address: the difference between AI systems that can \emph{dynamically retrieve} context from a user's personal or organizational knowledge corpus and those that require \emph{manual context attachment} by the user at each query. We term this the \textbf{Context Access Divide (CAD)}. For knowledge-intensive white-collar workers whose effective intellectual capital is distributed across tens of thousands of files, the CAD is not merely a convenience gap---it constitutes a qualitative threshold in AI usefulness. Below this threshold, the cognitive burden of context curation falls on the human, reproducing the very inefficiencies AI is supposed to eliminate. Because this interaction-level architectural difference systematically advantages certain classes of workers and organizations, its distributional consequences aggregate to the societal level, making it a macro-consequential micro-variable that complements \citeauthor{Sharp2025}'s person-level analysis. We propose \textbf{contextuality}---the degree to which an AI system autonomously accesses a user's accumulated knowledge capital---as a dimension of AI-mediated inequality that operates alongside, but at a different analytical level from, the availability, quality, and quantity dimensions. We formalize this divide by proposing a probabilistic model of human recall limits applied to context curation, grounded in the fan effect literature in cognitive psychology, and demonstrate formally how manual context attachment leads to a combinatorial collapse in task-success probability as corpus size and task conjunctivity grow---a collapse from which dynamic context retrieval architectures are structurally insulated. We examine the technical basis of this divide in the Model Context Protocol (MCP) and retrieval-augmented generation (RAG) architectures, and analyze its implications for the stratification of knowledge work.

\end{abstract}

\section{Introduction}

The emergence of autonomous AI agents---systems capable of planning, tool use, and multi-step task execution---marks a qualitative shift beyond conversational AI chatbots. As \citet{Sharp2025} observe, this shift is not merely technological but distributional: who can access capable agents, and at what level of capability, will shape the distribution of power and opportunity in economic and social life.

The \citet{Sharp2025} framework identifies three core dimensions along which agentic inequality manifests: \emph{availability} (whether one can access any agent at all), \emph{quality} (the capability level of the agent one can access), and \emph{quantity} (the number of agents one can deploy simultaneously). This framework represents an important conceptual advance over earlier digital divide scholarship, which focused on binary access to infrastructure.

Yet the framework, as currently formulated, operates at a level of analysis that may obscure a structurally significant divide operating within the ``availability'' and ``quality'' dimensions. Two users may both have nominal access to the same AI platform---the same subscription tier, the same underlying model---and yet experience qualitatively different AI utility depending on a single architectural feature: whether the system can \emph{autonomously retrieve} relevant context from the user's existing knowledge corpus, or whether the user must \emph{manually identify and attach} that context at each interaction.

We term this the \textbf{Context Access Divide (CAD)}. Unlike the three dimensions in \citet{Sharp2025}---which are person- and organization-level variables characterizing who can access agents---the CAD is an \emph{interaction-architecture-level} variable: a property of how each AI session is structured, specifically who bears the burden of context curation at the moment of each query. Yet because this micro-level architectural difference systematically advantages knowledge workers with large corpora and technical sophistication, its distributional consequences aggregate upward to the societal level. The CAD thus functions as a macro-consequential micro-variable, and we propose \textbf{contextuality} as a dimension of AI-mediated inequality that complements, but is not reducible to, the \citet{Sharp2025} framework. The CAD is particularly consequential for knowledge-intensive workers---researchers, legal professionals, consultants, writers, analysts---whose intellectual capital is distributed across large, heterogeneous collections of personal documents, notes, and files accumulated over professional careers. For such workers, the question of whether AI can dynamically access this corpus is not a marginal ergonomic concern; it determines whether AI functions as a genuine cognitive partner or merely as a sophisticated search interface that still requires human mediation.

It is worth pausing on why the CAD requires a different kind of attention than the inequality dimensions most readily recognized in public and scholarly discourse about AI. Discussions of AI-related inequality typically default to a familiar template inherited from older debates about the digital divide: some people can afford or access a resource and others cannot, and the resulting gap in capability tracks existing axes of disadvantage. This template is intuitive because it maps directly onto established sociological vocabulary---who has access, who does not---and requires no revision to how inequality is normally conceived. The CAD does not fit this template, and readers accustomed to it may reasonably wonder how a difference in interaction architecture, rather than in access per se, constitutes a form of inequality at all. An analogy familiar to the scholarly community itself may help. Consider two researchers who each have exactly the same number of hours allocated to research in a given week---an identical endowment of the resource that "availability" and "quantity" in the Sharp et al.\ sense would measure. The first researcher works alone: before any thinking can begin, she must search the literature herself, locate the relevant passages in her own accumulated notes and files, retrieve and often physically photocopy materials, and assemble these into raw material for thought. The second researcher works within a well-resourced research university, supported by research assistants and administrative staff who continuously organize, index, and retrieve from the literature and from the researcher's own archive; the researcher need only say "bring me this, and that," and the relevant materials appear. Both researchers have identical research time in the sense the conventional framework would measure, and both may have identical access to the same journals, databases, and computing resources. Yet the substantive research output the two can produce in that time differs enormously, because the cognitive labor of constituting the material basis for thought has been distributed differently between researcher and support staff. This is, of course, a familiar and consequential difference between research environments---it tracks, among other things, persistent disparities between well-resourced research universities and institutions where faculty must perform this labor unaided. The CAD names the AI-mediated analogue of this difference: a gap that access-based measures do not register because they ask whether one has an AI system, not how the cognitive labor of using it well is divided between the system and the person.

This paper proceeds as follows. Section~\ref{sec:background} situates the CAD within the existing literature on agentic inequality, digital divides, and knowledge work. Section~\ref{sec:technical} provides a technical account of the architectures that produce the CAD. Section~\ref{sec:dimension} develops the CAD as an analytical dimension, characterizing its structure and the threshold effects it generates. Section~\ref{sec:social} examines the social implications of the CAD for knowledge-work stratification. Section~\ref{sec:discussion} discusses limitations and directions for future research.

\section{Background and Related Work}
\label{sec:background}

\subsection{Agentic Inequality}

\citet{Sharp2025} define agentic inequality as ``disparities in power, opportunity, and outcomes arising from unequal access to, and capabilities of, AI agents'' (p.~1). The paper argues that AI agents differ from earlier digital tools in a crucial respect: they function as \emph{autonomous delegates} rather than passive tools, capable of pursuing complex goals independently. This delegation capacity introduces novel power asymmetries, including the ability to scale goal-pursuit far beyond individual human capacity and to engage in direct agent-to-agent competition.

The three-dimensional framework (availability, quality, quantity) provides a useful analytical structure. Its three dimensions operate at the level of \textbf{persons and organizations}: they characterize who has access to agents, at what capability level, and in what numbers. This is an important and appropriate level of analysis for questions about macro-distributional consequences---the kinds of systemic inequality that manifest in labor markets, income distributions, and political power. Within the ``quality'' dimension, the paper identifies factors such as core model capability, processing speed, and access to proprietary data feeds as differentiators. However, it does not examine how inequality can also emerge at a finer level of analysis: the \textbf{individual interaction}, where two users with nominally equivalent agent access may nonetheless experience radically different AI utility depending on a single architectural feature---whether the system autonomously retrieves context or requires the user to supply it manually.

This distinction between levels of analysis is important. The CAD is not a macro-distributional variable in the \citet{Sharp2025} sense; it is an \emph{interaction-architecture} variable---a property of how each AI session is structured, specifically who bears the burden of context curation at the moment of each query. Yet this micro-level architectural difference has macro-level consequences: because it systematically advantages knowledge workers with large corpora and technical sophistication, it reproduces and amplifies the same kinds of distributional inequalities that the \citet{Sharp2025} framework is designed to capture. The CAD thus connects micro-level interaction design to macro-level stratification outcomes, and it does so through a mechanism that existing frameworks do not name.

\subsection{Digital Divides and Knowledge Work}

Scholarship on digital divides has evolved from first-generation analyses of infrastructure access \citep{VanDijk2006} through second-generation attention to skills and usage \citep{Warschauer2003,Hargittai2002} toward third-generation analyses of outcomes and benefits \citep{Ragnedda2013}. A parallel evolution is evident in analyses of technology and knowledge work: from concerns about whether workers have access to information technology, to how AI affects the quality and distribution of cognitive labor.

Recent empirical work suggests that generative AI tools have heterogeneous productivity effects across workers, with larger gains accruing to less experienced workers in some domains \citep{Brynjolfsson2025,DellAcqua2026} but disproportionate benefits to senior workers in others \citep{HosseiniMaasoum2025}. This heterogeneity underscores the importance of analyzing not only whether workers use AI, but \emph{how} they use it---including what context they can bring to bear.

Recent work has also proposed extensions of the knowledge gap hypothesis to generative AI contexts: \citet{Morisco2026} argues that generative AI shifts the locus of inequality from information access to the critical evaluation of AI-generated outputs, with more educated users better positioned to question and contextualize what AI systems produce. The CAD analysis complements this perspective by identifying a structural condition that precedes the evaluation stage: whether the AI system can autonomously access a user's accumulated knowledge corpus in the first place. These are distinct but related mechanisms through which AI can either amplify or attenuate existing inequalities.

\subsection{Retrieval-Augmented Generation and MCP}

The technical literature on retrieval-augmented generation (RAG) establishes that LLM utility for knowledge-intensive tasks depends critically on the quality and scope of context supplied at inference time \citep{Lewis2020,Gao2024}. Early RAG systems required users or system designers to pre-define document collections and retrieval pipelines. More recent agentic RAG systems embed retrieval decisions into the model's reasoning flow, enabling autonomous determination of what to retrieve and when \citep{Singh2025}.

The Model Context Protocol (MCP), released by Anthropic in November 2024, provides a standardized interface through which AI agents can access external data sources, tools, and services without bespoke integration for each source. MCP enables AI agents to \emph{proactively query} external systems---including personal file repositories, knowledge bases, and organizational data stores---as part of normal conversational flow, rather than requiring users to manually retrieve and supply relevant materials.

The speed of MCP's adoption illustrates the magnitude of the integration problem it addresses. Server downloads grew from approximately 100,000 at launch in November 2024 to over 8 million by April 2025 \citep{ByteBridge2026}. The protocol was adopted by OpenAI (March 2025), Google DeepMind (April 2025), and Microsoft (May 2025). In December 2025, Anthropic donated MCP governance to the Agentic AI Foundation under the Linux Foundation, co-founded with Block and OpenAI, cementing its status as vendor-neutral infrastructure \citep{LinuxFoundation2025}. By Q1 2026, the protocol had surpassed 97 million monthly SDK downloads---an approximately 970-fold increase in thirteen months---and more than 10,000 active public MCP servers were in production use \citep{Anthropic2025,Knak2026}. An independent census from Q1 2026 indexed 17,468 MCP servers across registries \citep{Knak2026}. The protocol's trajectory from proprietary tool to open infrastructure in under eighteen months is without close recent precedent among AI standards.

However, platform-level MCP support does not automatically translate into DCRM for individual users. The gap between a platform \emph{supporting} MCP and a user \emph{experiencing} dynamic context retrieval involves additional layers of configuration, technical expertise, and organizational deployment---a distinction that is central to the CAD analysis developed in this paper.

It is important to note that dynamic context retrieval is not exclusive to any single AI provider or protocol. Open DCRM via MCP is available across multiple platforms including Claude (Anthropic), ChatGPT (OpenAI), and Gemini (Google DeepMind), all of which adopted MCP support by mid-2025. Walled DCRM---autonomous retrieval within a proprietary ecosystem---is implemented by Google (Gemini within Google Drive, Gmail, and Google Calendar), Microsoft (Copilot within Microsoft 365 and OneDrive), and others. The CAD framework is therefore not a comparison between specific commercial products but an analysis of architectural categories that cut across providers: the same provider may offer MAM in one interface and Walled or Open DCRM in another, and the same user may occupy different positions on the CAD spectrum depending on which interface and configuration they employ.

\section{The Technical Basis of the Context Access Divide}
\label{sec:technical}

\subsection{Three Architectures of Context Supply}

Contemporary AI assistant interfaces can be characterized along a spectrum defined by who bears primary responsibility for context curation, and how far that retrieval extends across data sources:

\medskip
\noindent\textbf{Manual Attachment Model (MAM).} The user is responsible for identifying which files, documents, or data are relevant to a given query and supplying them via explicit attachment mechanisms (e.g., file upload buttons, copy-paste). The AI system processes only what is explicitly provided. This model is common in consumer AI interfaces where cloud storage integrations are available, but the integration surfaces only as a mechanism for users to select and attach individual files---the selection judgment remains with the user.

\medskip
\noindent\textbf{Walled Dynamic Context Retrieval Model (Walled DCRM).} The AI system can autonomously retrieve context from data sources within a defined ecosystem---for example, a user's files, email, and calendar within a single provider's platform---without requiring manual attachment at each turn. The system determines what to retrieve and integrates it into the response autonomously, but this capability is bounded by the provider's ecosystem boundaries. Representative examples include the Gemini assistant embedded in Google Drive (which can search across Google Drive, Gmail, and Google Calendar dynamically) and Microsoft 365 Copilot (which retrieves across Office documents, Teams, and Outlook). Critically, Walled DCRM operates through proprietary, closed integrations: while these platforms may nominally support open protocols such as MCP, the dynamic retrieval capability within the walled environment does not extend to external data sources or third-party services. This architecture creates platform lock-in as a structural feature: users whose knowledge corpora are distributed across multiple providers, or who rely on locally stored files or specialized databases outside the provider's ecosystem, remain partially in MAM conditions even while experiencing DCRM within the walled environment.

\medskip
\noindent\textbf{Open Dynamic Context Retrieval Model (Open DCRM).} The AI system can autonomously query data sources across ecosystem boundaries---including personal knowledge bases, local file systems, email, calendars, specialized databases, and third-party services---as part of task processing. The system determines what context is relevant, retrieves it, and incorporates it without explicit user instruction at each turn. This architecture is enabled by open, vendor-neutral protocols such as MCP combined with agentic RAG pipelines and tool-use capabilities, allowing users to connect heterogeneous data sources regardless of provider.

\medskip
These three models are not simply points on a convenience spectrum; they represent qualitatively different distributions of cognitive labor between human and machine, and qualitatively different relationships between users and platform providers. The distinction between Walled DCRM and Open DCRM is particularly consequential: both relieve the user of manual attachment within their respective scopes, but only Open DCRM does so without imposing ecosystem dependency as the price of that relief.

\subsection{The Cognitive Burden of Context Curation}

Under the MAM, effective AI use requires the user to perform a prior retrieval task before the primary task can begin. This prior retrieval involves:

\begin{enumerate}[leftmargin=*]
\item \textbf{Recall}: The user must remember that a relevant document exists.
\item \textbf{Identification}: The user must locate the specific file among potentially thousands.
\item \textbf{Selection}: The user must judge which of multiple potentially relevant files to supply.
\item \textbf{Attachment}: The user must perform the mechanical act of uploading or referencing the file.
\end{enumerate}

Each of these steps imposes cognitive and temporal costs. More importantly, steps (1) and (2) introduce a structural limitation: the user can only supply context they already know exists and can locate. As personal knowledge corpora grow---in knowledge-intensive professional domains, corpora of tens of thousands of files are not unusual---the probability that a user will fail to recall or locate relevant material increases substantially. This pattern is consistent with findings from the personal information management literature: empirical diary studies of information re-finding find that the dominant cause of failed retrieval is not deficiencies in search tools but lapses in retrospective memory---the failure to recall that a relevant item exists at all \citep{Elsweiler2007}. Under the MAM, therefore, AI utility is bounded by the user's own retrieval capacity, which is itself bounded by human memory limitations.

Walled DCRM eliminates these costs within the provider's ecosystem, but a structurally analogous limitation applies at the ecosystem boundary: the user can only draw on context that resides within the walled environment. For knowledge workers whose intellectual capital is distributed across local files, multiple cloud providers, specialized databases, or domain-specific tools, the walled boundary reproduces the MAM's structural limitation at a higher level of aggregation. The user who stores research notes in a local knowledge base, working documents in one cloud provider, and correspondence across multiple email systems will find that Walled DCRM provides partial relief while leaving substantial portions of their corpus inaccessible to autonomous retrieval.

Under the Open DCRM, these costs are transferred to the AI system across ecosystem boundaries. The human specifies a goal; the system determines what context is needed and retrieves it from whatever sources are connected, regardless of provider. This transfer is not merely a time-saving convenience---it qualitatively alters the class of tasks for which AI assistance is effective, particularly for knowledge workers with large, heterogeneous, cross-ecosystem corpora.

\subsection{A Threshold Effect and Its Nested Structure}

We argue that the transitions across the three architectures exhibit threshold dynamics rather than a smooth gradient of improvement. The reason is that the utility of AI assistance is a nonlinear function of context completeness. For tasks requiring synthesis across multiple documents or reference to material the user has not recently consulted---particularly those characterized by high conjunctive context dependency, as elaborated below---the MAM may produce near-zero marginal utility even for technically sophisticated users, not because the AI model is incapable, but because the relevant context never enters the interaction.

The theoretical basis for expecting threshold dynamics rather than continuous improvement rests on a property of knowledge synthesis tasks: they are often characterized by what might be called \emph{conjunctive context dependency}, where the quality of the output depends not on the average relevance of supplied context but on the presence of specific critical elements. A legal brief that incorporates eight of ten relevant precedents but omits the two most on-point ones may be worse than useless---it may be affirmatively misleading. A literature review that covers the main findings in a field but misses the recent paper that overturns a key assumption will lead the researcher in the wrong direction. In such tasks, partial context does not yield proportionally partial utility; it yields a qualitatively different---and potentially inferior---output.

\subsubsection{A Formal Model of the MAM Floor Constraint}
\label{sec:formal-model}

We formalize this intuition with a simple probabilistic model. Consider a knowledge-synthesis task that requires $k$ specific, conjunctively necessary documents to be present in the AI system's working context for the output to be reliable---that is, the task fails (or degrades qualitatively) unless \emph{all} $k$ critical documents are supplied, not merely some subset.

Under the MAM, the user must recall and manually attach each critical document. We model the user's per-document recall probability as a function of corpus size $N$. The functional form is motivated by the \emph{fan effect}, a well-established and repeatedly replicated finding in cognitive psychology: as the number of items associated with a given retrieval cue increases, both the speed and the accuracy of retrieving any one specific associated item decline \citep{Anderson1974,AndersonReder1999}. In Anderson's original formulation and its extensions, retrieval accuracy and latency for a target item are governed by the \emph{associative fan}---the number of competing items sharing a retrieval cue---with greater fan producing systematically lower recognition accuracy and slower retrieval, an effect attributed to the division of limited associative activation across a growing number of competing memory traces \citep{Anderson1974,Schneider2012}. Corpus size $N$ in our model plays the role of associative fan: a knowledge worker's intention to retrieve a specific document is a retrieval cue (e.g., ``the paper on X'' or ``the note about Y''), and as the number of documents that could plausibly match that cue grows, retrieval accuracy and speed for the specific target document decline by the same mechanism documented in the fan effect literature. We capture this qualitative relationship, established at the level of mechanism though not previously parameterized for personal document corpora of the scale relevant here, with a logistic-decay function:
\begin{equation}
q(N) = q_{\min} + \frac{q_{\max} - q_{\min}}{1 + (N/N_0)^{\beta}}
\label{eq:q_recall}
\end{equation}
where $q_{\max}$ is the recall probability for a single document in a small corpus, $q_{\min}$ is the asymptotic floor recall probability retained even in very large corpora (reflecting that some documents remain salient or recently used regardless of corpus size), $N_0$ is the corpus size at which recall has decayed halfway between $q_{\max}$ and $q_{\min}$, and $\beta$ governs the steepness of the decay.

Assuming the $k$ critical documents must each be independently recalled and attached, the probability of MAM success is:
\begin{equation}
P_{\mathrm{MAM}}(\mathrm{success} \mid N, k) = q(N)^{k}
\label{eq:p_mam}
\end{equation}

This simple multiplicative structure already generates the threshold dynamics central to our argument: because $q(N) < 1$, success probability decays \emph{exponentially} in $k$, and because $q(N)$ itself decays in $N$, the combination produces a rapid collapse in MAM reliability as either corpus size or task conjunctivity increases. With illustrative parameters ($q_{\max}=0.95$, $q_{\min}=0.05$, $N_0=50$, $\beta=1$), a task requiring $k=3$ conjunctive documents has a MAM success probability of $0.51$ at a small corpus of $N=50$ documents, but falls to $0.0008$ at $N=1{,}000$ and to approximately $0.0001$ at $N=10{,}000$---a corpus size not unusual for an established knowledge worker's accumulated notes, correspondence, and reference materials. We emphasize that these specific parameter values are illustrative rather than empirically estimated; the qualitative conclusion---that MAM success probability collapses combinatorially as corpus size and task conjunctivity both increase---is robust to substantial variation in parameter choice, since it follows from the multiplicative structure of Equation~\ref{eq:p_mam} rather than from any particular calibration.

We extend the model to Walled and Open DCRM by modifying how retrieval probability depends on $N$. Under Walled DCRM, a fraction $\alpha \in [0,1]$ of the user's corpus resides within the provider's ecosystem and is retrieved autonomously with high probability $q_{\text{eco}}$; the remaining fraction $(1-\alpha)$ is retrieved only via the user's MAM-level recall $q(N)$. The effective per-document retrieval probability is then $q_{\text{eco}} \cdot \alpha + (1-\alpha) \cdot q(N)$, raised to the power $k$. Under Open DCRM, retrieval is performed by the system across the full corpus via agentic search rather than human recall, so retrieval probability $q_{\text{dcrm}}$ is high and approximately independent of $N$: $P_{\mathrm{OpenDCRM}}(\mathrm{success} \mid k) \approx q_{\text{dcrm}}^{k}$.

\begin{figure}[t]
\centering
\includegraphics[width=\textwidth]{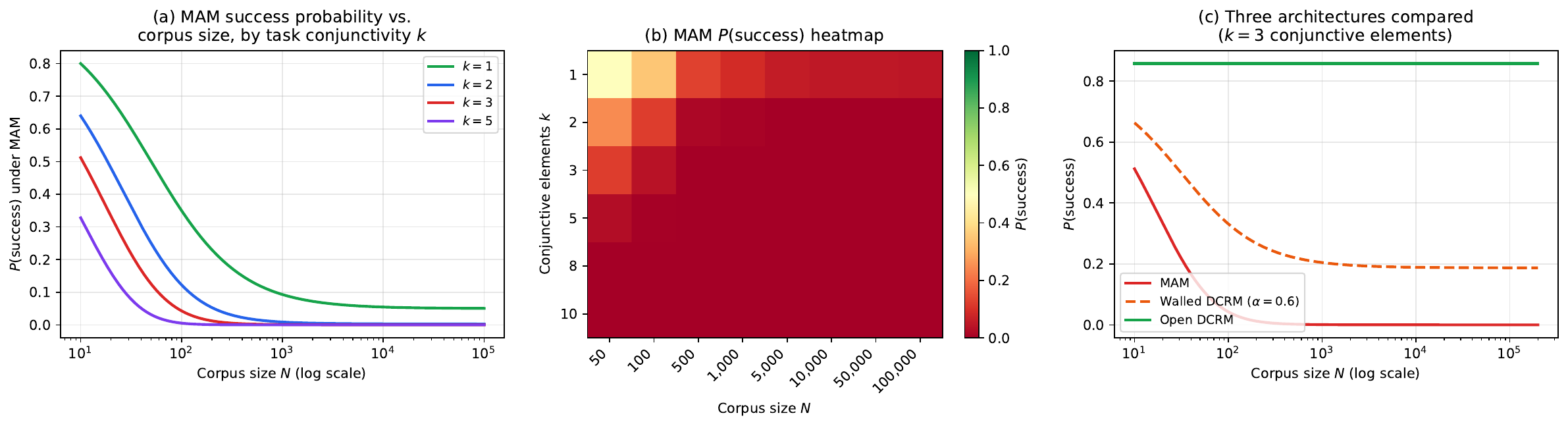}
\caption{Simulated success probability under conjunctive context dependency. (a) MAM success probability as a function of corpus size $N$, for varying levels of task conjunctivity $k$. (b) Heatmap of MAM success probability jointly across $N$ and $k$. (c) Comparison of the three architectures (MAM, Walled DCRM with $\alpha=0.6$, Open DCRM) at fixed conjunctivity $k=3$. Parameters: $q_{\max}=0.95$, $q_{\min}=0.05$, $N_0=50$, $\beta=1$, $q_{\text{eco}}=0.92$, $q_{\text{dcrm}}=0.95$. All parameter values are illustrative; see Section~\ref{sec:formal-model} for the underlying model and Section~\ref{sec:sim-limitations} for limitations.}
\label{fig:simulation}
\end{figure}

Figure~\ref{fig:simulation} plots these relationships. Panel (a) shows MAM success probability collapsing as corpus size grows, with the collapse accelerating sharply as task conjunctivity $k$ increases. Panel (b) presents this jointly as a heatmap across corpus size and conjunctivity. Panel (c) compares the three architectures directly at $k=3$: with illustrative parameters ($\alpha=0.6$, $q_{\text{eco}}=0.92$, $q_{\text{dcrm}}=0.95$), Open DCRM maintains success probability near $0.86$ regardless of corpus size, Walled DCRM stabilizes at an intermediate plateau around $0.19$ (reflecting the residual MAM-level failure on the $(1-\alpha)$ fraction of the corpus outside the ecosystem), and MAM collapses toward zero. At $N=10{,}000$, the model implies that Open DCRM is approximately $5{,}300$ times more likely to succeed at this conjunctive task than MAM. While the precise multiple is an artifact of the chosen parameters, the qualitative pattern---an architecture-dependent gap that widens by orders of magnitude as corpus size grows---is the structural claim the model is designed to illustrate, and it directly substantiates the nested-threshold argument developed below. The logistic-decay form of Equation~\ref{eq:q_recall} was chosen for analytical convenience; Appendix~\ref{app:robustness} verifies that the paper's qualitative conclusions are not an artifact of this choice by comparing it against an exponential-decay alternative motivated by an independent literature on large-pool semantic memory search.

\subsubsection{Limitations of the Formal Model}
\label{sec:sim-limitations}

This model is a theoretical illustration, not an empirical estimate, and several of its simplifying assumptions deserve explicit note. No single source of evidence available in the existing literature simultaneously satisfies all three conditions an ideal calibration would require---the correct retrieval mechanism, the correct scale, and the correct independent variable---so the case for the model's functional form rests on triangulation across three partial sources, each of which supplies one or two of these conditions while lacking the others, rather than on direct measurement. First, the parameters governing $q(N)$ ($q_{\max}$, $q_{\min}$, $N_0$, $\beta$) are chosen for illustrative plausibility rather than fitted to observed recall data in knowledge-work settings. The fan effect literature establishes the qualitative and mechanistic basis for $q(N)$'s decline---that retrieval accuracy for a cued item degrades as the number of competing associates grows---but fan effect experiments are typically conducted with small, controlled stimulus sets (fans of 1 to 4 or 5 associates per cue), not with corpora of thousands of personal documents, so the specific functional form and parameter values in Equation~\ref{eq:q_recall} extrapolate the established mechanism well beyond the range in which it has been directly measured. This extrapolation is a genuine limitation: we cannot rule out that the relationship between corpus size and recall probability saturates, accelerates, or otherwise departs from the logistic form at the scale of thousands or tens of thousands of documents relevant to professional knowledge work. This finding is complemented by diary-study evidence from the personal information management (PIM) literature: \citet{Elsweiler2007} found in a one-week diary study of 25 participants that the majority of failures to re-find previously encountered personal information were attributable to retrospective memory lapses---failing to recall that a specific item existed or where it was located---rather than to deficiencies in the retrieval system itself. Together, the fan effect literature supplies the cognitive mechanism and the PIM literature supplies ecological validity for the claim that recall failure, not search-tool inadequacy, is the operative bottleneck under MAM; neither, however, provides parameter estimates for $q(N)$ at the corpus scales central to this paper.

We note, moreover, that the absence of such estimates may not be a contingent gap awaiting future research so much as a structural feature of the phenomenon itself. Controlled fan effect experiments require participants to explicitly learn a stimulus set before retrieval is tested, and this learning phase imposes a practical ceiling on feasible set size: experiments with fans beyond single digits become infeasible within a standard laboratory session of one to a few hours. A personal knowledge corpus of the scale relevant to this paper---thousands to tens of thousands of documents accumulated over years of professional practice---lies many orders of magnitude beyond what any feasible controlled-encoding experiment could test within an ordinary experimental session, a session whose duration is itself comparable to a single block of knowledge work. On this reading, the methodological inaccessibility of large-corpus recall experimentation is not merely a limitation of the present model but suggestive corroborating evidence for the paper's central claim: if the task of manually identifying and supplying the right context from a corpus of this scale is too cognitively heavy to even be experimentally simulated within a timeframe comparable to ordinary knowledge work, this is consistent with---though it does not on its own demonstrate---the claim that the task is also too heavy to be reliably performed within the time constraints of that work in practice. We offer this as a suggestive interpretation rather than a formal argument, since the inaccessibility of an experimental paradigm has multiple possible explanations (cost, ethical constraints on prolonged sessions, lack of researcher interest in this scale) and is not by itself a measurement of task difficulty. Nonetheless, it is consistent with, and in our view modestly reinforces, the paper's broader argument that the cognitive demands of MAM at realistic corpus scales are not a minor inconvenience but a load-bearing constraint on what unaided human recall can accomplish.

Obtaining direct parameter estimates for $q(N)$ at the corpus scales relevant to professional knowledge work---for instance, through naturalistic or quasi-experimental designs that exploit researchers' or professionals' existing, already-encoded personal corpora rather than requiring fresh laboratory encoding---is an important direction for future empirical research, and one that the methodological reasoning above suggests may require departing from the controlled-encoding paradigm that has dominated fan effect research to date.

A further point requires clarification, since the cognitive memory literature does contain large-scale studies that might appear, at first glance, to contradict the model's central premise. \citet{Standing1973} famously demonstrated that recognition memory for pictures is "almost limitless": after a single exposure to up to 10,000 images, participants retained the ability to judge "have I seen this before?" with high accuracy, and the absolute number of items retained continued to increase with set size rather than collapsing. This finding, and the broader massive-memory literature it inaugurated, might seem to suggest that human memory capacity does not impose the kind of corpus-size penalty our model assumes. The apparent tension is resolved by a basic distinction in memory research between \emph{recognition} and \emph{cued recall}, which test fundamentally different retrieval processes. Recognition requires only a familiarity judgment given the target item itself as a cue (\emph{"is this the document I am thinking of?"}); cued recall requires generating or locating the target from a different, indirect cue without the target being presented (\emph{"which document discusses topic X?"}). The MAM scenario this paper models is a cued-recall problem: the knowledge worker does not have the target document in front of them for a yes/no judgment, but must generate its identity and location from a topical or task-based cue. This is precisely the condition under which the fan effect and cue-overload phenomena operate, and under which performance is well documented to degrade as the number of items sharing a cue increases \citep{Watkins1975}---a degradation that does not arise, or arises far less severely, under recognition testing. The massive-memory literature is therefore not in tension with the model; it characterizes a different retrieval operation than the one MAM places on the user. We flag this distinction explicitly because the cued-recall/recognition contrast, while foundational in cognitive psychology, is not always observed in computer science and information-retrieval treatments of human memory, where "memory capacity" claims are sometimes invoked without specifying which retrieval operation is at issue. (Appendix~\ref{app:robustness} notes an analogous caveat regarding the exponential-decay robustness check.)

Second, the model assumes independence across the $k$ critical documents' recall probabilities, which may not hold if, for example, documents on the same topic tend to be recalled or forgotten together (positive dependence), or if recalling one document serves as a retrieval cue for related documents (also positive dependence, which would attenuate the collapse we model). Third, the model treats $k$ as fixed and known, whereas in practice a user under MAM does not know in advance how many or which documents are conjunctively necessary---arguably making the real-world MAM condition worse than the model suggests, since the user cannot verify success even when it occurs. Fourth, we model only recall failure, not the inverse error of attaching irrelevant documents that dilute the AI system's effective context window; this is a related but distinct mechanism that could be incorporated in future extensions. Despite these simplifications, the qualitative conclusion central to our argument---that success probability under MAM degrades combinatorially with corpus size and task conjunctivity, while DCRM architectures are structurally insulated from this degradation---follows directly from the multiplicative form of Equation~\ref{eq:p_mam} and is robust to the specific parameter values chosen.

The reader may reasonably ask why a mechanism validated only at small fan sizes is extended to corpora several orders of magnitude larger. Our answer is that we treat this extension as an explicit extrapolation rather than a validated measurement, and address the resulting uncertainty through three complementary moves rather than one: triangulation across mechanistically, ecologically, and scale-relevant but individually incomplete sources of evidence (above); an argument that direct experimental validation at the relevant scale is not merely unperformed but methodologically difficult to perform within the controlled-encoding paradigms available (below); and a demonstration that the paper's qualitative conclusions are insensitive to the specific functional form chosen for $q(N)$ (Appendix~\ref{app:robustness}). None of these moves substitutes for direct empirical estimation, which we identify as a priority for future work; together, they are intended to establish that the model's qualitative behavior is a reasonable theoretical conjecture warranting empirical test, not an artifact of an arbitrarily chosen curve.

Walled DCRM removes this floor constraint within its ecosystem, producing a qualitative improvement over MAM for users whose corpora are primarily housed within that ecosystem. However, it introduces a new threshold: for users with cross-ecosystem corpora, the walled boundary becomes the binding constraint, and the effective utility of Walled DCRM approximates MAM for the out-of-ecosystem portions of their corpus.

Open DCRM removes both constraints, enabling autonomous retrieval across the full scope of connected sources. The result is a nested threshold structure: MAM $\rightarrow$ Walled DCRM $\rightarrow$ Open DCRM represents two qualitative discontinuities, not a single one, with the practical significance of each discontinuity varying by the user's corpus distribution across ecosystems.

This nested structure implies that the CAD creates multiple qualitative discontinuities in AI usefulness. Users positioned at different points in this structure are not simply experiencing better or worse AI assistance; they are experiencing different categories of cognitive partnership with AI systems---and the category boundary that matters most for a given user depends on where their knowledge capital is stored.

An important boundary condition follows from this last point. For users whose knowledge corpora are substantially consolidated within a single provider's ecosystem---for example, a professional whose documents, email, and calendar all reside within Google Workspace---the discontinuity between Walled DCRM and Open DCRM may be negligible in practice, because the walled boundary does not bisect their working corpus. For such users, Walled DCRM may deliver nearly the full benefit that Open DCRM would provide. The CAD's distributional consequences are therefore conditional on the degree of corpus heterogeneity: the more a worker's intellectual capital is distributed across ecosystems, the more consequential the Walled-to-Open DCRM threshold becomes. This conditionality does not undermine the CAD analysis but refines it. It predicts that the workers most disadvantaged by the absence of Open DCRM are precisely those whose professional practices involve cross-ecosystem knowledge management---a pattern that, as we argue in Section~\ref{sec:social}, is characteristic of senior knowledge workers with accumulated, heterogeneous corpora spanning multiple tools, institutions, and career stages.

\section{The Context Access Divide as a Complementary Dimension}
\label{sec:dimension}

\subsection{Structural Position within the Sharp et al.\ Framework}

The CAD occupies a distinctive structural position relative to the \citet{Sharp2025} framework. As noted in Section~\ref{sec:background}, the three existing dimensions---availability, quality, and quantity---are \textbf{person- and organization-level} variables: they characterize the distribution of agent access across actors. The CAD, by contrast, is an \textbf{interaction-architecture-level} variable: it characterizes a property of how each individual AI session is structured, specifically whether context retrieval is borne by the human or the system at the moment of each query.

This difference in level of analysis does not render the CAD orthogonal to the \citet{Sharp2025} dimensions---it is partially correlated with all three. It is partially related to ``quality'': a more capable agent might be expected to have better context retrieval, but the relationship is imperfect. A user may have access to a high-capability AI model (high quality) but interface with it through a platform that implements only MAM (low CAD position). Conversely, a technically sophisticated user may configure DCRM capabilities on top of a mid-tier model, achieving high CAD position without access to frontier model quality. It is partially related to ``availability'' insofar as MCP-enabled DCRM requires both platform support and user technical capacity to configure---but availability in the \citet{Sharp2025} sense refers to agent access per se, not to any particular architectural feature of how that agent operates. And it is partially related to ``quantity'' insofar as organizations deploying fleets of agents can more easily invest in DCRM infrastructure than individual users.

The key contribution of identifying CAD as a distinct dimension is precisely that these partial correlations are imperfect. Two workers with identical scores on availability, quality, and quantity may nonetheless face a qualitative discontinuity in AI usefulness that depends entirely on interaction-level architecture. This gap is invisible to the \citet{Sharp2025} framework as currently formulated, because that framework does not descend to the level of individual interaction design.

We therefore propose \textbf{contextuality} as a dimension of AI-mediated inequality: the degree to which an AI system can autonomously access and integrate the user's accumulated knowledge capital without requiring human mediation at each interaction. Though contextuality operates at the interaction level, its distributional consequences aggregate to the person and societal level---making it a macro-consequential micro-variable that complements \citeauthor{Sharp2025}'s person-level framework without being reducible to it. The relationship to the existing three dimensions is one of cross-level articulation rather than parallel addition: contextuality names the mechanism by which interaction-architecture choices translate into the distributional outcomes that availability, quality, and quantity describe.

\subsection{Dimensions of the CAD}

Within the contextuality dimension, we identify several sub-components:

\begin{itemize}[leftmargin=*]
\item \textbf{Corpus scope}: What types of data sources can the system access? (Local files, cloud storage, email, specialized databases, real-time data)

\item \textbf{Retrieval autonomy}: Does the system determine what to retrieve, or does the user specify? (Full autonomy vs.\ user-directed vs.\ hybrid)

\item \textbf{Corpus scale}: How large a knowledge corpus can the system effectively retrieve from? (Bounded by index capacity, context window limits, and retrieval precision)

\item \textbf{Temporal currency}: Can the system access recently created or modified materials without manual synchronization?

\item \textbf{Integration friction}: What technical expertise and configuration effort is required to achieve DCRM? (API configuration, MCP server setup, index maintenance)
\end{itemize}

These sub-components determine not only whether a user experiences DCRM rather than MAM, but the quality of DCRM they experience.

\subsection{Who Is Most Affected?}

The CAD has heterogeneous impact across worker types. Its effects are most pronounced for workers who:

\begin{enumerate}[leftmargin=*]
\item Maintain large personal knowledge corpora accumulated over time (researchers, lawyers, consultants, writers, long-tenured professionals)
\item Engage in tasks requiring synthesis across multiple prior work products (literature review, case preparation, strategic analysis)
\item Work on projects where relevant prior material is not consistently at the top of mind (non-linear knowledge work)
\item Cannot feasibly pre-curate context for each AI interaction due to time constraints or corpus scale
\end{enumerate}

For workers whose tasks are primarily bounded and self-contained---where relevant context fits in a single conversation---the MAM/DCRM distinction has less practical significance. The CAD therefore exhibits an interaction effect with the nature of work: it is most consequential for the most cognitively complex knowledge work.

\section{Social Implications}
\label{sec:social}

\subsection{Stratification of Knowledge Work Benefit}

If the CAD creates a qualitative threshold in AI usefulness for complex knowledge work, and if access to DCRM is unevenly distributed, then the productivity benefits of AI adoption will be unevenly distributed in ways that existing agentic inequality frameworks do not capture.

Several mechanisms suggest that DCRM access is likely to be concentrated among workers who already possess advantages:

\medskip
\noindent\textbf{Organizational resources.} Enterprises with IT departments and AI deployment budgets are better positioned to configure MCP-enabled DCRM systems for their employees than individuals or small organizations. Early enterprise deployments demonstrate productivity gains that confirm the value of DCRM: Block, an early adopter and major contributor to the MCP ecosystem that co-founded the Agentic AI Foundation with Anthropic and OpenAI, reported that since deploying its Goose AI agent (MCP-connected) internally, engineers shipped over 40\% more production code per engineer, with some risk-underwriting models that previously took a full quarter to build completed in a fraction of the time \citep{Block2026,Fortune2026}. These gains are available primarily to employees of organizations with the resources and technical expertise to deploy MCP infrastructure.

\medskip
\noindent\textbf{Technical capital.} Configuring DCRM---setting up MCP servers, maintaining indices, managing authentication---currently requires non-trivial technical expertise. The Stacklok State of MCP in Software 2026 report found that even among senior technical leaders in software companies, only 41\% had achieved some form of production MCP server deployment \citep{Stacklok2026}---and production MCP deployment encompasses a broad range of use cases beyond personal knowledge retrieval, suggesting that DCRM adoption for the purpose of dynamic context access from personal corpora is likely lower still. Among non-technical knowledge workers, the barrier is substantially higher. This creates a skill-based access barrier correlated with education and professional background.

\medskip
\noindent\textbf{Platform selection.} Different AI platforms implement MAM, Walled DCRM, or Open DCRM by design choice, not solely as a function of subscription cost. Users who are aware of and able to select Open DCRM-capable platforms gain advantage over otherwise equivalent users who are unaware of the architectural distinction. A further complication is that Walled DCRM, while representing a genuine improvement over MAM, simultaneously functions as a lock-in mechanism: users who migrate their knowledge corpora into a single provider's ecosystem to achieve Walled DCRM gain contextuality within that ecosystem at the cost of cross-ecosystem portability. Providers such as Google and Microsoft have structural incentives to offer capable Walled DCRM precisely because it deepens ecosystem dependency, potentially displacing Open DCRM as the dominant architecture even as MCP and similar open protocols mature. Users capable of recognizing and resisting this substitution---by maintaining cross-ecosystem corpora and investing in Open DCRM configuration---are disproportionately technically sophisticated and institutionally resourced.

\medskip
\noindent\textbf{Corpus quality.} DCRM amplifies the value of existing knowledge corpora. Workers with larger, better-organized, more extensively annotated personal knowledge bases benefit more from DCRM than workers with smaller or less structured corpora. This creates a compounding dynamic in which prior investment in knowledge organization yields increasing returns as DCRM becomes available.

\subsection{The ``AI Productivity Premium'' and Knowledge Work Stratification}

\citet{Zhang2026} identifies an ``AI productivity premium''---a widening gap between researchers with AI access and skills and those without---as itself a stratification phenomenon warranting social scientific analysis. The CAD analysis suggests that this premium has internal structure: even among AI-using knowledge workers, those with DCRM access may experience substantially larger productivity gains than those limited to MAM, particularly in domains characterized by large accumulated corpora and extended, non-linear project timelines.

This implies that the appropriate unit of analysis for AI-driven knowledge-work stratification is not simply ``AI user vs.\ non-user'' but includes finer distinctions in the architecture of AI-human interaction. Survey instruments and empirical studies that treat ``AI use'' as a binary variable, or that fail to distinguish MAM from DCRM users, will systematically underestimate the variance in AI productivity effects.

\subsection{Compounding Effects and the Risk of Entrenchment}

The compounding dynamics noted above---DCRM amplifying the value of existing knowledge corpora, which in turn were built through prior professional investment---suggest a risk of entrenchment. Workers who have accumulated large knowledge corpora and can access DCRM may experience productivity gains that accelerate their knowledge accumulation further, widening the gap with workers who cannot. This is structurally analogous to the ``Matthew effect'' in science \citep{Merton1968}---cumulative advantage accruing to those already advantaged---but operating at the level of AI-mediated knowledge work.

The compounding is bidirectional: DCRM users accumulate knowledge faster, and the larger corpus they accumulate yields greater DCRM benefit. This positive feedback loop has no natural equilibrium and may produce increasing stratification over time, even if the technical barrier to DCRM declines.

\subsection{Platform Governance Implications}

The CAD analysis has implications for AI platform governance that extend beyond those discussed in the agentic inequality literature. If the choice to implement MAM rather than DCRM in a platform is a design decision that systematically disadvantages users with large knowledge corpora, it raises questions about whether such choices should be transparent to users and whether regulatory frameworks for AI should encompass context retrieval architecture.

The current situation---in which nominally equivalent AI access (same subscription tier, same model) can mask substantial differences in effective utility depending on context retrieval architecture---is consistent with the ``pacing problem'' identified by \citet{Sharp2025}: institutional adaptation has not kept pace with the distributional consequences of technical design choices. The rapid normalization of MCP as infrastructure (its donation to the Linux Foundation in December 2025, backed by Anthropic, OpenAI, Google, Microsoft, and others) may accelerate platform convergence on DCRM support, but the \emph{configuration gap}---the difference between a platform supporting MCP and a user experiencing DCRM---will persist as a governance challenge.

A structurally important pattern in the current platform landscape is what we term the \textbf{Walled DCRM substitution dynamic}: major platform providers offer capable Walled DCRM---autonomous retrieval within a proprietary ecosystem---as a functional substitute for Open DCRM, while simultaneously maintaining ecosystem boundaries that prevent cross-provider retrieval. This dynamic is illustrated by Microsoft's integration of OpenAI models into Microsoft 365 Copilot and OneDrive, and Google's deployment of Gemini within Google Drive, Gmail, and Google Calendar. Both implementations provide genuine DCRM functionality within their respective ecosystems, but are designed as closed, proprietary integrations that deepen platform dependency rather than enabling cross-ecosystem portability.

The Microsoft/OpenAI case is instructive for governance analysis. OpenAI was founded with a stated mission of ensuring that artificial general intelligence benefits ``all of humanity.'' Yet Microsoft's investment of over \$13 billion since 2019, combined with exclusive cloud deployment arrangements, has resulted in OpenAI's frontier models being deployed primarily within Microsoft's Walled DCRM ecosystem. Whether or not this arrangement is consistent with OpenAI's founding rationale, it illustrates a broader structural dynamic: the tension between open access to AI capabilities and the commercial incentives of platform providers to channel those capabilities through proprietary ecosystems. This tension is a governance challenge that existing AI regulation frameworks have not yet addressed.

This pattern suggests a broader dynamic: providers have structural incentives to offer Walled DCRM precisely because it combines the user-experience benefits of dynamic retrieval with the commercial benefits of ecosystem lock-in. Open DCRM, which would deliver contextuality without lock-in, does not generate equivalent commercial returns for any single provider, creating a systematic market under-provision of the architecture that would most benefit users with heterogeneous, cross-ecosystem corpora.

This market structure analysis suggests that governance interventions may need to go beyond transparency requirements to address the structural incentives that favor Walled over Open DCRM. Relevant policy instruments might include interoperability mandates requiring platforms to support open context retrieval protocols, data portability requirements ensuring users can migrate knowledge corpora across providers without loss of DCRM functionality, and procurement standards that favor Open DCRM architectures in public sector AI deployment. The broader political-economic dimensions of this dynamic---including its relationship to historical patterns of enclosure in academic knowledge infrastructure such as the commercialization of open statistical software and the acquisition of open scholarly repositories by commercial publishers---raise questions of distributive and reciprocal justice that warrant dedicated analysis beyond the scope of this paper.

\section{Discussion and Limitations}
\label{sec:discussion}

\subsection{Limitations}

Several limitations should be noted. First, this paper is primarily conceptual; empirical measurement of the CAD and its distributional effects remains an important agenda item. Measuring the productivity effects of MAM vs.\ DCRM in ecologically valid knowledge-work settings would require controlled studies with workers possessing large personal corpora---a methodologically challenging requirement.

Second, the boundaries among the three architectures identified in this paper---MAM, Walled DCRM, and Open DCRM---are not always sharp in practice, and the Walled DCRM category in particular encompasses a wide range of implementations differing in ecosystem scope, retrieval autonomy, and openness to third-party extension. Future analytical work may need to develop more fine-grained measures of the contextuality dimension that can capture this variation.

Third, the CAD analysis is predicated on a relatively stable distinction between the retrieval models. This distinction may evolve as AI platforms converge on DCRM capabilities. However, even if DCRM becomes universal, the CAD analysis would remain relevant for understanding \emph{transition dynamics}---the period during which the divide exists will have distributional consequences that may persist after the technical gap closes.

\subsection{Future Directions}

We identify three priority directions for future work. First, empirical studies measuring the productivity effects of context retrieval architecture across knowledge-worker populations differing in corpus size and work type. Second, analysis of the determinants of DCRM adoption---which organizations, professional domains, and user populations achieve DCRM access, and through what pathways. Third, governance analysis of context retrieval architecture as a dimension of AI platform regulation, including transparency requirements and potential design mandates.

The CAD also suggests a research agenda at the intersection of personal knowledge management (PKM) scholarship \citep{Grundspenkis2007,Smedley2009} and AI systems design: how can individuals and organizations build knowledge corpora that are maximally compatible with DCRM architectures, and what are the equity implications of differential capacity to do so?

\section{Conclusion}

The framework introduced by \citet{Sharp2025} is an important contribution to the analysis of AI-driven inequality. Their three dimensions---availability, quality, and quantity---provide a powerful vocabulary for describing who can access AI agents and at what capability level. This paper has argued that a complementary analysis is needed at a different level: the individual interaction, where the architecture of context retrieval determines whether AI can serve as a genuine cognitive partner or merely a sophisticated text interface requiring constant human mediation.

The Context Access Divide names this interaction-level variable, and \textbf{contextuality} is our proposed label for the dimension it instantiates. The CAD does not simply add a fourth item to the \citet{Sharp2025} list; it identifies a mechanism operating at a different analytical level whose distributional consequences aggregate upward to produce the kinds of inequality that \citeauthor{Sharp2025} are concerned with. In this sense, contextuality is a cross-level complement to the existing framework, not a parallel extension of it.

The CAD is not a marginal ergonomic difference. For knowledge-intensive workers whose professional capital is distributed across large accumulated corpora, it represents a qualitative threshold in AI usefulness. The threshold dynamics of the CAD mean that users on opposite sides are not simply experiencing better or worse AI---they are experiencing different categories of cognitive partnership with AI systems.

As AI agents become more deeply integrated into knowledge work, the architectural choices that determine context retrieval will increasingly shape the distribution of AI productivity benefits. The rapid adoption of MCP as open infrastructure---from a novel protocol in November 2024 to a Linux Foundation standard backed by all major AI providers by December 2025---demonstrates that the technical capacity for universal DCRM is already being built. Whether this capacity translates into equitable access, or instead compounds existing advantages among technically sophisticated, well-resourced knowledge workers, will depend on choices in platform design, organizational deployment, and governance that are not yet well understood. Ensuring that these choices are made transparently, and that the resulting inequality is recognized and addressed, requires analytical frameworks that attend to both levels of analysis---and to the mechanisms that connect them.

\appendix

\section{Robustness Check: Alternative Functional Form for $q(N)$}
\label{app:robustness}

The logistic-decay form of Equation~\ref{eq:q_recall} (Section~\ref{sec:formal-model}) was chosen for analytical convenience, and a natural question is whether the paper's qualitative conclusions depend on this particular choice. An independent line of cognitive research offers an alternative functional form against which to check robustness. Studies of semantic memory search using the verbal fluency paradigm---in which participants generate as many exemplars as possible from a single category cue (e.g., "name as many animals as you can"), drawing on a natural vocabulary of tens of thousands of words---model the declining rate of successful retrieval over the course of search using an \emph{exponential} decay function rather than a logistic one \citep{Rohrer1995}. In this paradigm, the cue is fixed and search proceeds through an effectively large, naturalistic item pool, making it structurally closer to the corpus-scale retrieval problem this paper addresses than are the small-fan laboratory paradigms discussed in Section~\ref{sec:formal-model}, even though the relevant independent variable in Rohrer et al.'s model is search duration within a fixed retrieval episode rather than corpus or vocabulary size per se. Its use here therefore borrows a functional form with independent theoretical motivation in a structurally adjacent large-pool cued-retrieval paradigm, not a direct empirical estimate of $q(N)$.

\begin{figure}[h]
\centering
\includegraphics[width=\textwidth]{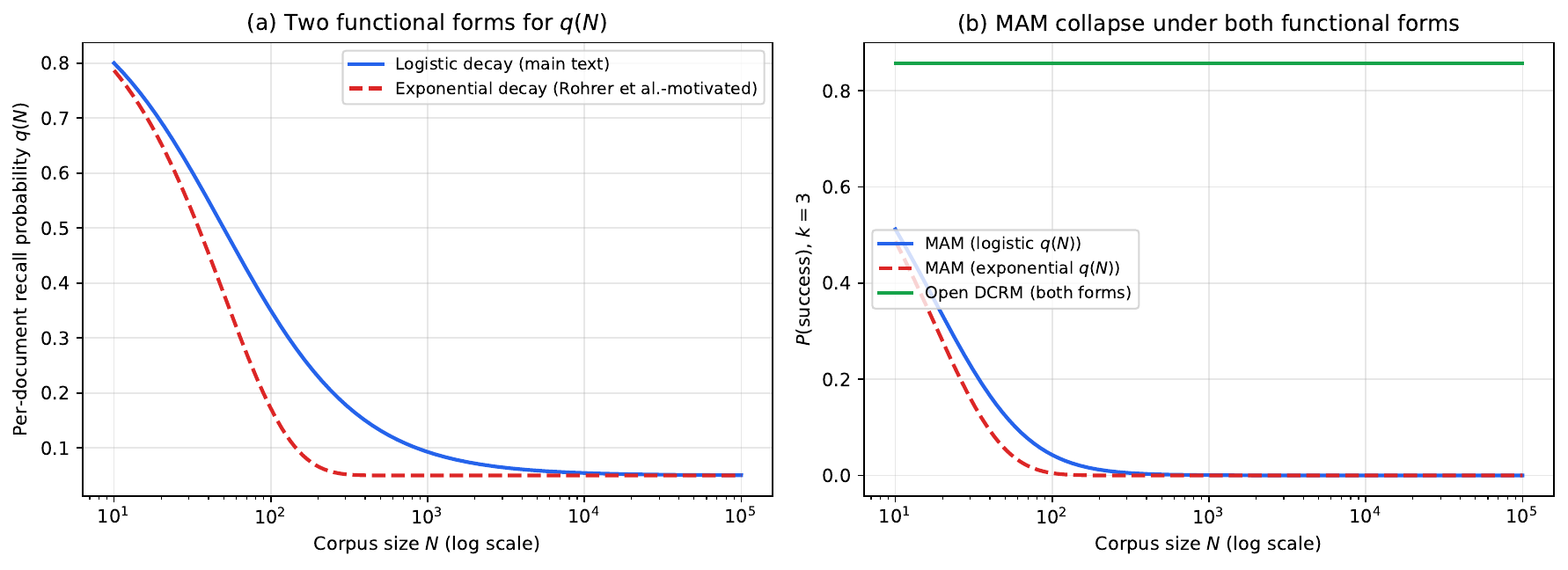}
\caption{Robustness of the MAM collapse to functional form. (a) The logistic decay form used in the main text (Equation~\ref{eq:q_recall}) compared against an exponential decay form motivated by Rohrer et al.'s (1995) model of semantic fluency search. (b) MAM success probability under both forms at $k=3$, compared against Open DCRM. The qualitative collapse pattern is preserved across both functional specifications.}
\label{fig:robustness}
\end{figure}

Motivated by this work, we define an alternative exponential-decay form, $q_{\exp}(N) = q_{\min} + (q_{\max} - q_{\min}) \cdot e^{-N/N_0}$, and recompute $P_{\mathrm{MAM}}(\mathrm{success})$ under this specification, holding all other parameters fixed. Figure~\ref{fig:robustness} compares the two functional forms directly. The exponential form produces a qualitatively similar---in fact, somewhat steeper---collapse: at $N=100$, $k=3$, MAM success probability is $0.043$ under the logistic form versus $0.005$ under the exponential form; by $N=1{,}000$, both forms have collapsed to well below $0.001$. The Open-DCRM-to-MAM advantage ratio at $N=10{,}000$ is $5{,}303\times$ under the logistic form and $6{,}859\times$ under the exponential form. The central qualitative claim of this paper---that MAM success probability collapses combinatorially with corpus size while DCRM architectures remain structurally insulated from this collapse---thus holds under both functional forms, and the exponential form, despite being motivated by an independent body of research on naturalistic, large-pool semantic retrieval, does not weaken the paper's argument; if anything, it suggests the logistic form may be conservative relative to at least one empirically grounded alternative.

\medskip
\noindent\textbf{Acknowledgments:} None.

\noindent\textbf{Conflicts of interest:} None.

\noindent\textbf{Data availability:} No new data were generated for this conceptual paper.

\bibliographystyle{plainnat}
\bibliography{cad_paper}

\end{document}